\documentclass[
 reprint,
 secnumarabic,
 amssymb,
 amsmath,
 apl,
 groupedaddress,
 twocolumn,
 preprint,onecolumn,
 frontmatterverbose,
 a4paper
]{revtex4}
\usepackage{graphicx}
\usepackage{bm}
\usepackage{dcolumn}
\usepackage[colorlinks=true,linkcolor=blue,citecolor=blue]{hyperref}%
\newcommand{\textOmega}{$\Omega$}\renewcommand{\textmu}{$\mu$}
\expandafter\ifx\csname package@font\endcsname\relax\else
 \expandafter\expandafter
 \expandafter\usepackage
 \expandafter\expandafter
 \expandafter{\csname package@font\endcsname}%
\fi
\begin{document}


\title{Self-pulsing of a micro thin cathode discharge}

\author{A. Wollny, M. Gebhardt, T. Hemke, R.P. Brinkmann, and T. Mussenbrock}

\affiliation{Institute of Theoretical Electrical Engineering, Ruhr University Bochum, 44780 Bochum, Germany}

\begin{abstract}
Microplasmas operated at atmospheric pressure show a number of peculiar dynamic phenomena. One of these phenomena is self-pulsing, which is characterized by intrinsic pulsing behavior of a DC driven plasma discharge. This work focuses on the numerical simulation of self-pulsing in a micro thin cathode discharge operated in atmospheric pressure argon. By means of a hybrid plasma model we show self-pulsing of the discharge in the expected MHz frequency range and described its actual origin.
\end{abstract}

\maketitle




Microplasmas operated at atmospheric pressure have gained increasing attention since Schoenbach and co-workers did their pioneering work in the mid of the 1990th \cite{Schoenbach1996}. Microplasmas show strong nonequilibrium behavior although they are collisional \cite{Penache2002}. Electrons are hot (up to a few eV) whereas ions and atoms (and/or molecules) of the neutral gas are cold (about room temperature). Microplasmas are characterized by high electron densities (up to a few 10$^{17}$ cm$^{-3}$) and high power densities (up to some 100 kW/cm$^{3}$). Due to this extraordinary properties microplasmas have found widespread of technological and biomedical applications. Examples can be found in the area of surface treatment, sterilization, and lighting \cite{Becker2006, Becker2010}.

\begin{figure}
\includegraphics[width=\columnwidth]{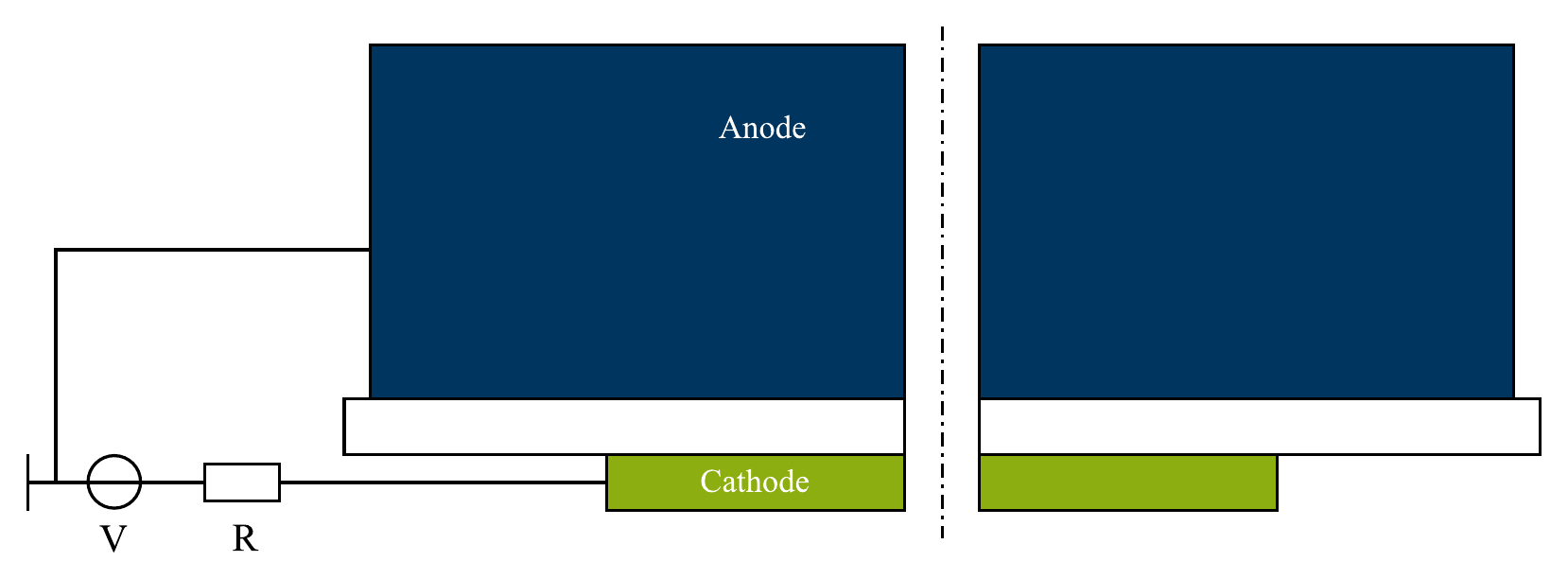}
\caption{Sketch of the $\mu$TCD with circuitry. Width of cathode and dielectric is 100~\textmu m, width of anode is 1~mm, hole diameter is 200~\textmu m. Applied voltage and series resistor are 1000~V and 25 k\textOmega, respectively.}
\label{fig:scetch}
\end{figure}

Microplasmas show unexpected dynamical behavior. One particular phenomenon that is experimentally observed in DC driven micro discharges is self-pulsing, which occurs as quasi periodic peaks in the discharge current. It is argued that this phenomenon can be understood as the periodic transition from the Townsend regime to the abnormal glow and vice versa. For a micro hollow cathode discharge operated at gas pressures between 50 and 100 Torr Rousseau and co-workers experimentally found the pulsing frequency ranging from a few tens to a few hundreds of kilohertz \cite{Rousseau2006, Aubert2007}. For the micro thin cathode discharge ($\mu$TCD) proposed by Czarnetzki and co-workers the self-pulsing phenomenon is in the megahertz range \cite{Du2011}.

Before the advent of microplasmas Petrovic and Phelps discussed certain oscillations of low current electrical discharges between plane-parallel electrodes \cite{Petro1993, Jelen1993, Phelps1993}. They propose lumped circuit models in order to describe the interaction of the nonlinear negative differential resistance ``behavior'' of the discharge with the external circuit. To study the self-pulsing phenomenon of micro discharges Hsu and Graves did pioneering work \cite{Hsu2003}. They also propose a lumped circuit model for the self-pulsing regime of micro hollow cathode discharges. Chabert et al. acted on the suggestions of Hsu and Graves and set up an improved lumped circuit model in order to carefully study the dynamics of the system \cite{Chabert2010}. Kolobov et al. described the transition from a Townsend discharge to a normal discharge using a two-dimensional numerical model \cite{Kolobov1994}. Their numerical model is based on a fluid description of electron and ion transport coupled with Poisson's equation, with the ionization source depending on the local field strength or provided by a Monte Carlo simulation of the fast electrons. The model is applied to an argon discharge, for a product of pressure and gap length in the 1-10 Torr cm range (which holds also for microplasmas). In 2003 Arslanbekov and Kolobov re-investigated the phenomenon by coupling two commercial simulation tools \cite{Arslan2003}. Low-current self-generated oscillations in a rectangular hollow cathode discharge in helium gas were investigated using a two-dimensional self-consistent hybrid model by Donk\'o \cite{Donko1999}. The model combines Monte Carlo simulation of the motion of fast electrons and a fluid description of slow electrons and positive ions. The low-frequency oscillations of approx. 20 kilohertz were found to arise as an effect of the interaction of the gas discharge and the external electric circuit consisting of a stable voltage source, a series resistor and a capacitor formed by the discharge electrodes.

In this work we study self-pulsing of the DC driven $\mu$TCD. The cross section of the $\mu$TCD is schematically depicted in Fig.~\ref{fig:scetch}. The difference between the micro hollow cathode discharge described by Schoenbach \cite{Schoenbach1996} and the $\mu$TCD investigated here is that the cathode is significantly thinner than the anode. The cathode has a thickness of 100 $\mu$m whereas the anode is 1000 $\mu$m thick. The dielectric has a thickness of 100 $\mu$m. The hole has a diameter of 200 $\mu$m.

\begin{figure}[t!]
\includegraphics[width=0.8\columnwidth]{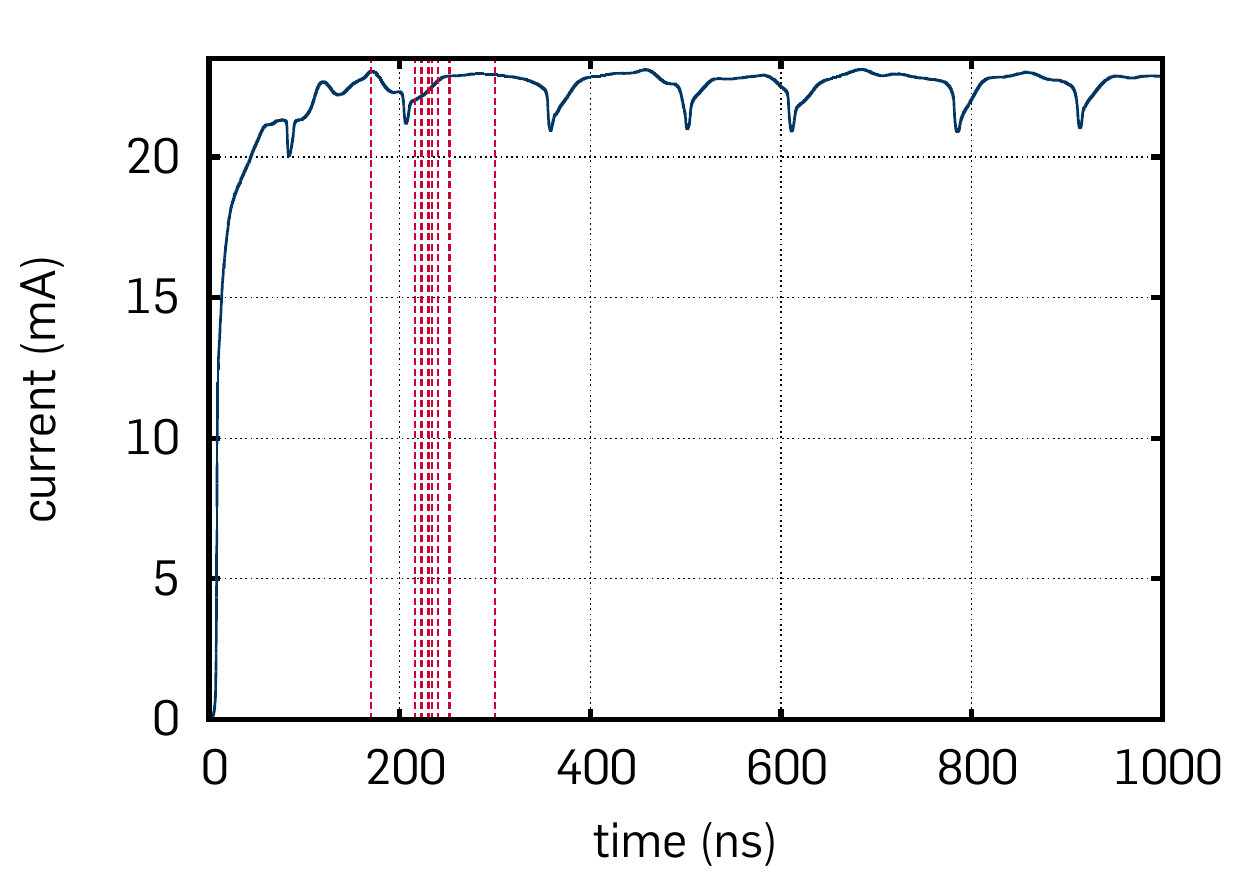}
\caption{The discharge current (mA) over time (ns). The vertical lines at 170~ns, 216~ns, 223~ns, 230~ns, 234~ns, 240~ns, 252~ns, and 300~ns correspond to electron density plots in Fig.~\ref{fig:ne}}
\label{fig:i}
\end{figure}

\begin{figure}[t!]
\includegraphics[width=0.8\columnwidth]{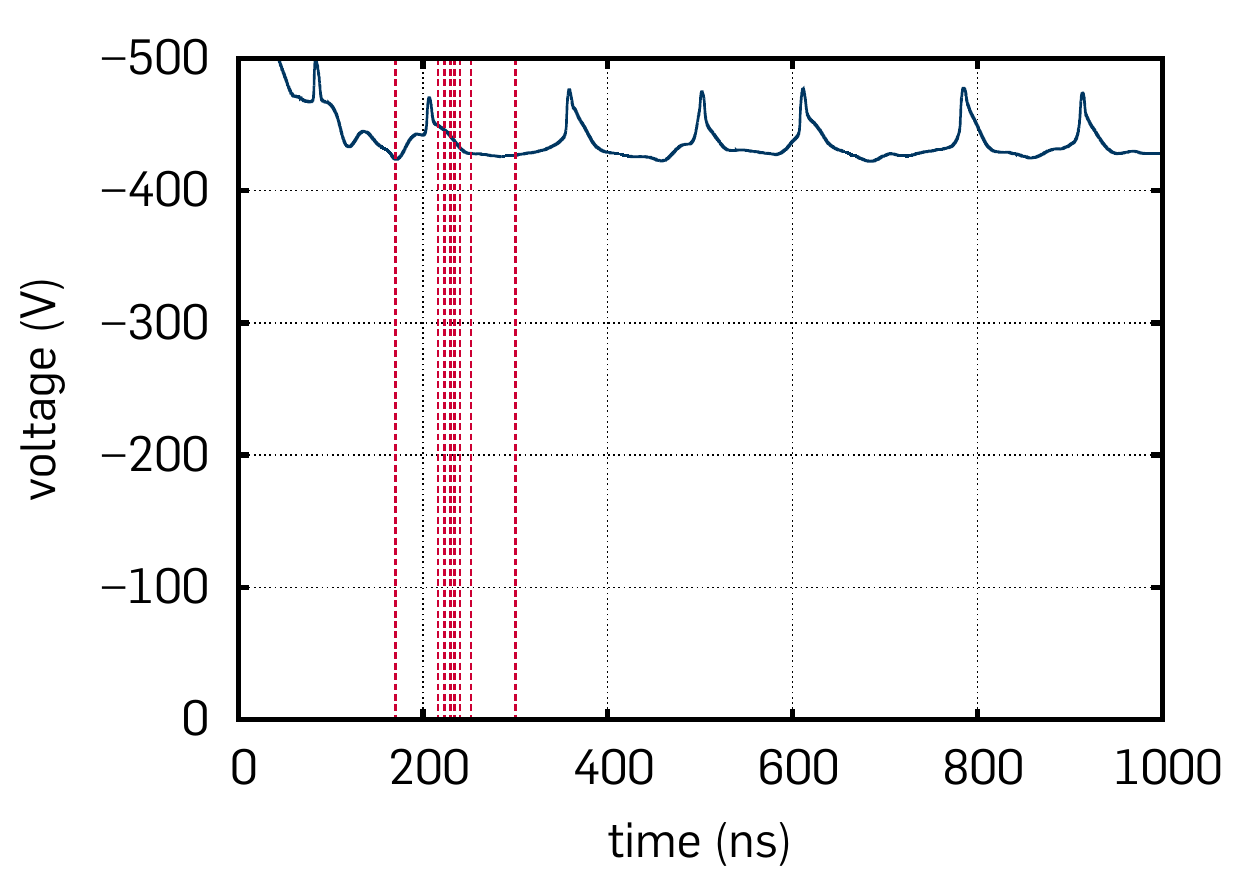}
\caption{The cathode voltage (V) over time (ns). The vertical lines at 170~ns, 216~ns, 223~ns, 230~ns, 234~ns, 240~ns, 252~ns, and 300~ns correspond to the electron density plots in Fig.~\ref{fig:ne}}
\label{fig:u}
\end{figure} 

In order to study the dynamics of the discharge we apply the two-dimensional hybrid plasma model \textit{nonPDPSIM} by Kushner and co-works, which is described in detail in \cite{Babaeva2007}. Here we briefly discuss the relevant physical equations. The electrical potential is calculated in electrostatic approximation giving Poisson's equation. Coupled to drift-diffusion equations charged species and a surface balance equation self-consistency is reached. The set of equations is simultaneously integrated in time using an implicit iteration scheme. Updates of densities and mean velocities of charged species are followed by an implicit update of electron temperature by solving the electron energy balance equation. To capture the non-Maxwellian behavior of electrons electron transport coefficients and rate coefficients are obtained by solving a zero-dimensional Boltzmann equation for electrons. Secondary electrons are treated in frame of a Monte-Carlo simulation. The discharge is sustained in argon at atmospheric pressure. The species in the model are electrons, Ar(3s),
Ar(4s), Ar(4p), Ar$^+$, Ar$^*_2$ and Ar$^+_2$. The reaction mechanism is summarized in Ref. \cite{Babaeva2007}.
Due to its the small thickness the cathode is easily heated by the plasma leading to thermionic electron emission from the cathode surface. This significantly contributes to power deposition into the discharge. However, this effect can clearly be separated from the self-pulsing  and is therefor not accounted for in this afford.


\begin{figure*}[t!]
\includegraphics[width=\textwidth]{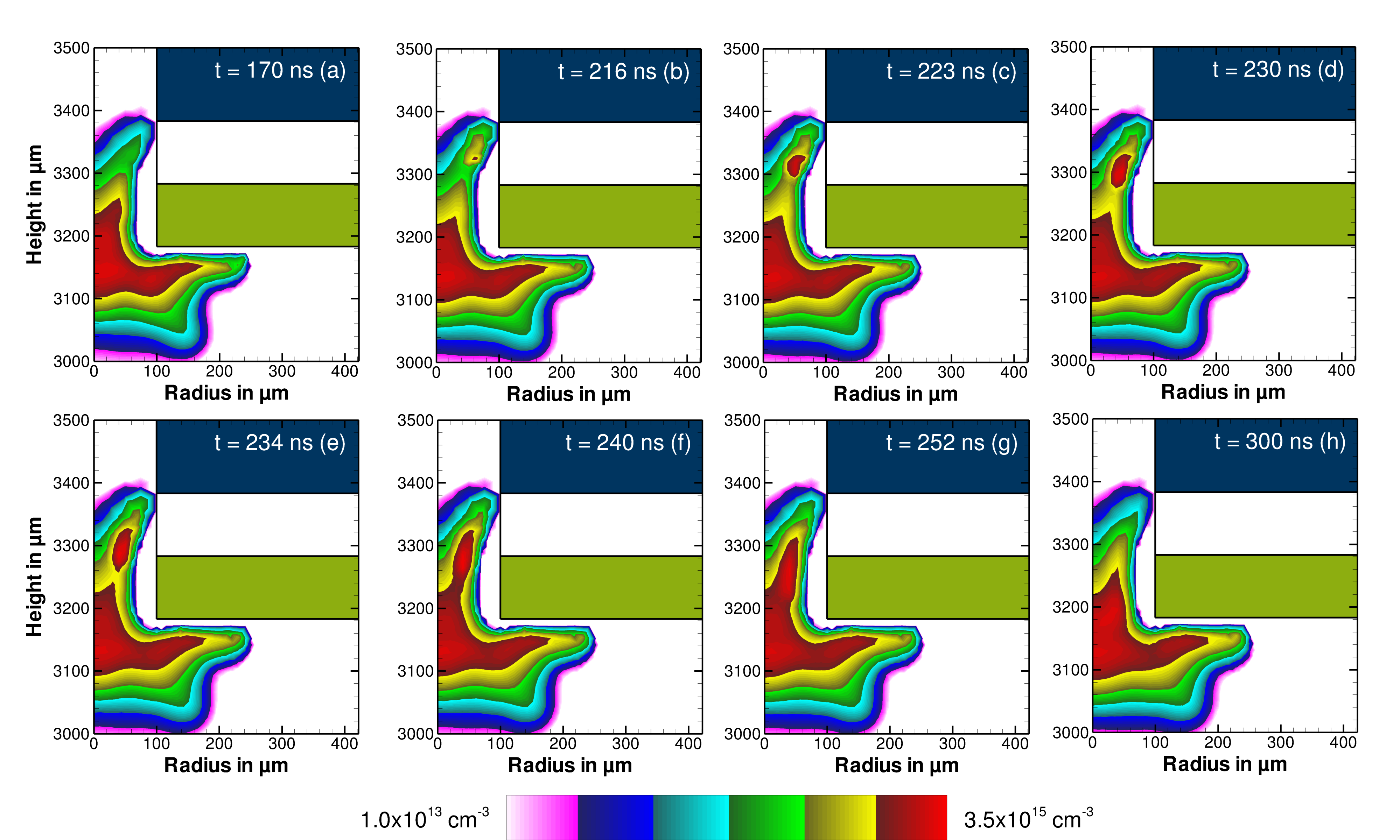}
\caption{Electron density in plasma region as function of time on log scale}
\label{fig:ne}
\end{figure*}

The simulation addresses a time interval of 900~ns and is performed on a computational domain of 500 \textmu m width and 1~mm height. A DC driving voltage of -1000~V is applied to the cathode via a series resistor of 25 k\textOmega. Electron emission is described applying a $\gamma$ coefficient of 0.15 for the cathode surface facing the plasma region.

The electron density as a function of time in the interval of 170~ns to 300~ns is presented in Fig.~\ref{fig:ne}. At the instant time of 170~ns the simulation has reached a quasi steady state with a flat maximum of about $2.5\times10^{15}$~cm$^{-3}$ at the edge of the cathode. A potential plot (not shown here) shows a equipotential area in the surrounding of this maximum, indicating a glow mode plasma. We like to call this area core in the following. In the following 60~ns a second maximum in the same order of magnitude is formed in the vicinity of the dielectric (Fig.~\ref{fig:ne}b,c). In contrast to the core this maximum has a peak shape. It appears that this density peak is the nucleus of a pulse. In Fig.~\ref{fig:ne}e the peak starts to advance towards the cathode and merges finally with the core. Another density peak forms approximately 40~ns later and the process is repeated. Interestingly the density peak launches from different positions along the dielectric resulting in a non-constant pulsing frequency.

Fig. 2 illustrates the electron and ion density distribution at the instant of time of $233$ ns, when the simulation reached a quasi steady state. The discharge shows a glow-like character with a flat maximum of about 2.5$\times$10$^{15}$ cm$^{-3}$ at the edge of the cathode. We obtain another maximum of the electron density of the same order in the vicinity of the dielectric. In contrast to the flat maximum at the cathode edge this maximum shows a peak shape. Looking at the dynamics of the plasma this maxima appears to be the beginning of a pulse. We focus on the phenomenon in Fig. 4. Here the temporal evolution of the inner maximum is given. The peak expands and moves in a beam towards the main maxima over a period of approx. $120$ ns and eventually vanishes. Another peak is formed approx. $30$ ns later and the whole process starts again. Interestingly in each period the ``beam'' launches from different positions along the dielectric and the pulsing frequency is not constant. Consequently, the discharge shows chaotic behavior in the given parameter regime. Fig. 3 shows the discharge current over a period of $1000$ ns. Here the self-pulsing behavior can clearly be seen as well.


The authors gratefully acknowledge valuable discussions with Dr. Natalia Babaeva, Dr. Zhongmin Xiong, and Prof. Mark Kushner from the University of Michigan at Ann Arbor. The authors also acknowledge financial support by the Deutsche Forschungsgemeinschaft in the frame of Research Group FOR1123 \emph{Physics of Microplasmas}.



%






\begin{thebibliography}{10}

\bibitem{Schoenbach1996} K.H. Schoenbach, R. Verhappen, T. Tessnow, F.E. Peterkin, W.W. Byszewski, Appl. Phys. Lett. 68, 13 (1996)

\bibitem{Penache2002} C. Penache, M. Miclea, A. Brauning-Demian, O. Hohn, S. Schossler, T. Jahnke, K. Niemax, and K. Schmidt-B\"ocking, Plasma Sources Sci. Technol. 11, 476 (2002)

\bibitem{Becker2006} K.H. Becker, K.H. Schoenbach, and J.G. Eden, J. Phys. D: Appl. Phys. 39, R55 (2006)

\bibitem{Becker2010} K.H. Becker, H. Kersten, J. Hopwood, J.L. Lopez, Eur. Phys. J. D 60, 437 (2010)

\bibitem{Rousseau2006} A. Rousseau, X. Aubert, J. Phys. D 39, 1619 (2006)

\bibitem{Aubert2007} X. Aubert, G. Bauville, J. Guillon, B. Lacour, V. Puech, and A. Rousseau, Plasma Sources Science and Technology 16, 23 (2007)

\bibitem{Du2011} B. Du, S. Mohr, D. Luggenh\"olscher, and U. Czarnetzki, J. Phys. D 44, 125204 (2011)

\bibitem{Petro1993} Z.Lj. Petrovi\'c and A.V. Phelps, Phys. Rev. E 47 2806 (1993)

\bibitem{Jelen1993} B.M Jelenkovi\'c, K. R\'ozsa K, and A.V. Phelps, Phys. Rev. E 47 2816 (1993)

\bibitem{Phelps1993} A.V. Phelps, Z.Lj. Petrovi\'c, and B.M. Jelenkovi\'c, Phys. Rev. E 47 2825 (1993)
 
\bibitem{Hsu2003} D.D. Hsu and D.B. Graves, Journal of Physics D: Applied Physics  36, 2898 (2003)

\bibitem{Chabert2010}  P. Chabert, C. Lazzaroni, and A Rousseau,  Journal of Applied Physics  108, 113307 (2010)
   
\bibitem{Kolobov1994} V.I. Kolobov and A. Fiala, Phys. Rev. E 50 3018 (1994)

\bibitem{Arslan2003} R.R. Arslanbekov and V.I. Kolobov, J. Phys. D: Appl. Phys. 36 2986 (2003)

\bibitem{Donko1999} Z. Donk\'o, J. Phys. D: Appl. Phys. 32, 1657 (1999)    

\bibitem{Kushner2004} M.J. Kushner, J. Appl. Phys. 95, 846 (2004)

\bibitem{Babaeva2007} N.Y. Babaeva, R. Arakoni, and M.J. Kushner, J. Appl. Phys. 101, 123306 (2007)

\end{thebibliography}
\end{document}